\documentclass{article}
\usepackage{amsfonts}
\usepackage{amssymb}
\usepackage{amsmath}
\usepackage{amsthm}
\usepackage{mathtools}
\begin{document}

\title{More on the fact that Rastall = GR}

\author{Alexey Golovnev\\
{\small {\it Centre for Theoretical Physics, the British University in Egypt,}}\\ 
{\small {\it BUE 11837, El Sherouk City, Cairo Governorate, Egypt}}\\
{\small agolovnev@yandex.ru}
}
\date{}

\maketitle

\begin{abstract}

Rastall gravity is the same as General Relativity, with a simple algebraic redefinition of what is called the energy-momentum tensor. Despite it having been very clearly explained by M. Visser several years go, there are still many papers claiming big differences between the two formulations of gravitational equations and trying to use them for problems of physics. When going this way, the totally ignored task is to explain why the conserved energy-momentum quantities and the quantities used for other purposes are different from each other. Moreover, when researchers are using the non-conserved energy density and pressure for determining the sound speed, it is just inconsistent with the Rastall gravity. I carefully explain all this, and also show how one could construct a variational principle for producing equations in the Rastall form.

\end{abstract}

\section{Introduction}

Modifying the gravity theories is very important, both for potential phenomenological applications and for better understanding of general relativity itself. The field of these attempts is very wide and diverse. Many options come about on our way, many ideas fade and start to develop anew, many directions exist for a very long time and get more and more modifications on top of their initial premise, and many once forgotten projects get revived at some point in time.

In particular, there was an odd idea of Rastall many years ago \cite{Rastall}, to start from denying the covariant "conservation" of the energy-momentum. It did not get too much attention at that time, however it has been relatively popular in the recent years again. In terms of the mathematical structures behind the theory of Einstein, it is an idea very difficult to implement naturally. The Levi-Civita covariant divergence of the gravity part of equations is identically zero, simply due to the action being a diffeomorphism-invariant object depending on the metric field only.

Therefore, getting an object with non-vanishing divergence in the pure gravity part already requires some new structures in it. At the same time, the description of matter must be changed, too. If we introduce the matter part by an action invariant under diffeomorphisms, then the energy-momentum tensor is on-shell conserved when defined in the usual way. Indeed, a variation coming from coordinate changes leaves the action identically invariant, therefore if the variation with respect to the matter fields is made to vanish by the equations of motion, then the variation with respect to the metric does have zero in its Levi-Civita divergence.

At the level of equations of motion, the (somewhat vaguely justified) idea of Rastall was to change the relative coefficient in front of $R_{\mu\nu}$ and $R g_{\mu\nu}$ terms in the Einstein tensor \cite{Rastall}. And as usual, we equate it to the energy-momentum tensor
\begin{equation}
\label{Rastall}
R_{\mu\nu} + \left(l - \frac12 \right) R g_{\mu\nu} = T_{\mu\nu},
\end{equation}
then as Visser correctly pointed out \cite{Visser}, it simply amounts to redefinition of what is called the energy and momentum, with no change to the gravity dynamics:
\begin{equation}
\label{RastEin}
G_{\mu\nu} = {\tilde T}_{\mu\nu} \qquad \mathrm{with} \qquad {\tilde T}_{\mu\nu}= T_{\mu\nu} + \frac{l}{1-4l}\cdot  T g_{\mu\nu}.
\end{equation}
Note that our $l$ is his $\frac{\lambda}{4}$.

Despite this very clear explanation, there are still many papers which appear to use Rastall reformulation of GR for doing physics. Because of that, this short note is devoted to supporting the Visser's viewpoint \cite{Visser} and uncovering some problems of works which go this way of modifying gravity. As a side remark, I will also show that it is actually not so difficult to construct an action for Rastall gravity, contrary to what is often said in the current literature.

\section{Modified description of matter}

As we have just mentioned, the Rastall equation (\ref{Rastall}) can be rewritten in a fully GR way (\ref{RastEin}). Vice versa, given GR with the energy momentum tesnor ${\tilde T}_{\mu\nu}$, we can write it down as a Rastall theory with a new definition of the tensor as
\begin{equation}
\label{back}
T_{\mu\nu} = {\tilde T}_{\mu\nu} - l \cdot {\tilde T} g_{\mu\nu}.
\end{equation}
Once more, this is the statement of Visser \cite{Visser}: there is no difference between GR and Rastall gravity but the redefinition of the energy-momentum tensor.

Then there were some claims that this fact means nothing since any gravity theory can be written in this way \cite{antiV}. This is of course true, and not only of a gravity theory. Any tensor-of-rank-two equation can be turned into this form, even without gravity at all. In the latter case, we would need to put the Einstein tensor also to the right hand side, making some ${\mathfrak T}_{\mu\nu}=0$ into $G_{\mu\nu}={\mathfrak T}_{\mu\nu}+G_{\mu\nu}$. This is an extreme case, however for many modified gravity theories rewriting them in this way does not go too far from it. For example, the $f(R)$ gravity, also mentioned in the Ref. \cite{antiV} in this context, will then get an important part of its gravitational dynamics, including the new scalar degree of freedom, into the effective energy-momentum tensor.

However, it is this objection \cite{antiV} to the Visser's claim \cite{Visser} what means nothing. The point is that, in the case of Rastall, what is effectively put to the right hand side of the Einstein equation (\ref{RastEin}) has nothing to do with gravity per se. There is no curvature, nor any other non-trivial geometric entity there. If we assume that $T^{\mu}_{\nu}$ describes some properties of pure matter, then so does the new object ${\tilde T}^{\mu}_{\nu}$ via (\ref{back})
$$T^{\mu}_{\nu} = {\tilde T}^{\mu}_{\nu} -  l \cdot {\tilde T} \delta^{\mu}_{\nu}.$$

Let's briefly consider an explicit example. For the sake of simplicity, and because it is the case of many papers applying Rastall gravity to physical problems, let's assume that the matter content is that of an ideal fluid. At a given point, we can always choose the normal coordinates in such a way that the metric is Minkowski and the connection coefficients are zero at this point and the fluid is at rest there. Then the redefinition (\ref{back}) can be read as changing the notions of the energy density and pressure as
\begin{eqnarray*}
\rho & =  & (1-l)\cdot {\tilde\rho} + 3l\cdot \tilde p,\\
p & = &  l\cdot \tilde\rho + (1-3l)\cdot {\tilde p}.
\end{eqnarray*}
And either inverting this transformation matrix or using the transformation law from the formula (\ref{RastEin}), one can get the inverse transforms:
\begin{eqnarray*}
\tilde\rho & =  & \frac{1-3l}{1-4l}\cdot \rho - \frac{3l}{1-4l}\cdot  p,\\
\tilde p & = & - \frac{l}{1-4l} \cdot \rho + \frac{1-l}{1-4l} \cdot  p,
\end{eqnarray*}
also given in the original work of Rastall \cite{Rastall}. 

Unfortunately, there are inaccuracies of the corresponding formulae in the Visser's work \cite{Visser}. Namely, he had lost the factors of $3$ in front of Einstein pressure in the expression of Rastall pressure and in front of Rastall energy density in that of Einstein energy density, as well as had confused the sign in front of the energy density in the former expression. Otherwise, I totally agree with him. And the main message is again that there is a very simple one-to-one correspondence between the Rastall quantities and those quantities which must obey the usual covariant conservation laws.

In other words, there does exist the covariantly conserved energy and momentum tensor ${\tilde T}_{\mu\nu}$ for the matter content. If the matter equations of motion allowed for some solutions with non-conservation of it, then those would be prohibited by the gravity equations of motion, most probably leading to the problem of an over-determined system of equations. Therefore, whatever matter content is assumed in the equations of Rastall gravity, it is either a pathological case of the two parts of the theory contradicting each other, or it does have all the usual conservation laws, precisely as in GR.

In other words, for the gravity and matter equations to be compatible with each other, the latter must also impose the covariant conservation of the energy-momentum tensor ${\tilde T}_{\mu\nu}$, or in terms of the Rastall quantities
\begin{equation}
\label{nuconsv}
\bigtriangledown_{\nu} T^{\nu}_{\mu} + \frac{l}{1-4l}\cdot \partial_{\mu} T =0,
\end{equation}
with non-conservation even in a flat spacetime except for the traceless theories like electrodynamics. It goes beyond the initial idea of Rastall \cite{Rastall} that the energy-momentum might depend on the curvature. However, he had never implemented such an idea, with all the later applications I know always taking some purely  matter contribution for the tensor $T_{\mu\nu}$. Formally, the idea was that the divergence of $T_{\mu\nu}$ be given by a gradient of the Ricci scalar, however due to all the very same equations (\ref{Rastall}) it can be rewritten in the form of Eq. (\ref{nuconsv}). It implies non-conservation even when the spacetime is assumed to be well-approximated by Minkowski metric, for otherwise the matter and the gravity equations of motion would generically contradict each other. If Rastall had really done what he was for, it would have rather been in the form of $f(R,{\mathcal L}_m)$ theories.

Mathematically, there is a one-to-one correspondence between the GR and Rastall theory (\ref{Rastall}) solutions, acknowledged even by Rastall himself \cite{Rastall}. One must simply change from one definition of the energy-momentum tensor to another (\ref{RastEin}). If some researchers declare a physical difference  \cite{antiV}, they have to look for it in the description of matter, not gravity. For those who are worried about the energy conditions, for example, this might entail an important difference, as long as they assume that it is the $T_{\mu\nu}$ tensor which has the meaning of the physical energy and momentum, and not the ${\tilde T}_{\mu\nu}$, even though I would say that the latter is the only reasonable option. 

All in all, barring possible internal inconsistencies, we have a matter content which does have a covariant conservation law for the quantity ${\tilde T}_{\mu\nu}$. The task for the proponents of Rastall gravity is then to explain why do we call another tensor $T_{\mu\nu}$ the energy-momentum, and why would we derive the physical properties from it. Referring to it as to the gravitational charge makes no good sense since it is simply due to our ad hoc rewriting of the Einstein equations. Therefore, the tensor ${\tilde T}_{\mu\nu}$ has the meaning of both being the standard gravitational charge for the Einstein tensor and representing the covariantly conserved quantities. If anything, there must be something in the physics of matter which could potentially justify any attention to another tensor, $T_{\mu\nu}$. One must then explain why they use the non-conserved tensor in cosmology for the sound speed \cite{cosmo} or the equation of state \cite{eqst} and in the physics of compact stars for the energy conditions \cite{Waleed} or the sound speed \cite{Waleed, Adel}.

\section{The uses of a wrong sound speed}

At the same time, it would be quite difficult to arrange for some gravitationally-important changes in the behaviour of matter without running into problems with a GR-equivalent model of gravity, such as the one by Rastall. And indeed, one of immediately obvious issues arises when many authors assume that, due to whatever mysterious a reason, the sound speed velocity is $c_s^2=\frac{\partial p}{\partial\rho}$, and then they make some physical conclusions from it. From stability and causality concerns we usually require that $0\leq c_s^2 \leq 1$, or even $0\leq c_s^2 \leq \frac13$. Of course, substituting the proper quantities by the Rastall ones surely changes these derivatives.

I won't go for discussing the meaning of these constraints on $c_s^2$. Even if we believe in causality, the superluminality does not necessarily mean a genuine causality violation \cite{BMV}. Moreover, the group velocity squared greater than one, $\frac{\partial p}{\partial\rho}>1$, does not necessarily imply any physical superluminality in the system at hand \cite{cause}. For that we would rather need to look at the characteristics, or the front velocity, while the usual interpretation of the group velocity requires that a wave packet is a substantially stable entity. Leaving all these interesting issues aside, I would like to show that the very idea that generically $c_s^2=\frac{\partial p}{\partial\rho}$ is simply wrong in Rastall gravity.

If we think in terms of simple sound waves in an ideal fluid, a fluid element changes its velocity due to the pressure gradient. It is almost a high school exercise to derive $c_s^2=\frac{\partial p}{\partial\rho}$ then, from the Newton laws and the energy under consideration being the kinetic energy of the fluid. It's easy indeed. However, it then contains the usual energy conservation as an important part of the theory. Therefore, there is no reason to believe that the same result must be true in other types of systems, too. Or we should be able to explain why it works with non-conserved physical quantities, instead of $\tilde \rho$ and $\tilde p$.

To state it again, it is a question about a modified theory of matter, why the sound speed is supposed to be given by the Rastall type quantities instead of the conserved ones which also do exist in the model. Moreover, remember that gravity equations themselves impose important restrictions on the behaviour of matter. If there is only one matter degree of freedom in the Universe, its motion is basically dictated by the covariant conservation. Therefore, one cannot think that the changed sound speed can be easily incorporated into the system with no problem, contrary to what was assumed in papers on cosmology \cite{cosmo} and pulsars \cite{Adel}.

Let's look at what happens with the theory of cosmological perturbations. In the conformal Newtonian gauge, the perturbed metric can be presented as
$$a^2 (t) \cdot \left((1+2\Phi)dt^2 - (1-2\Psi)d{\overrightarrow x}^2\right)$$
with the two Newtonian potentials $\Phi$ and $\Psi$. Then, assuming an ideal fluid in the right hand side, the off-diagonal spatial part of equation makes the potentials equal to each other, 
$$\Phi=\Psi.$$
Taking the rest of equations in the "Einstein frame" shape (\ref{RastEin}), we may use the standard results from the cosmological perturbation theory, though it is not a very difficult calculation anyway. Then the temporal component tells us that
\begin{equation}
\label{temp}
2\bigtriangleup \Phi - 6{\mathcal H}(\dot\Phi + {\mathcal H}\Phi) = a^2 \delta \tilde \rho
\end{equation}
while the diagonal spatial part takes the form of
\begin{equation}
\label{spat}
2\ddot\Phi + 6{\mathcal H} \dot\Phi +2(2 \dot{\mathcal H} + {\mathcal H}^2)\Phi = a^2 \delta \tilde p.
\end{equation}
In the "Rastall frame" (\ref{Rastall}) we would simply get two other linear combinations of these equations. One way or another, the two equations (\ref{temp},\ref{spat}) do imply together that
\begin{equation}
\label{wave}
\ddot\Phi + 3{\mathcal H} \left(1+\frac{\partial\tilde p}{\partial\tilde \rho}\right) \dot\Phi +\left(2 \dot{\mathcal H} + \left(1+3\frac{\partial\tilde p}{\partial\tilde \rho}\right) {\mathcal H}^2 \right)\Phi - \frac{\partial\tilde p}{\partial\tilde \rho} \bigtriangleup \Phi = 0.
\end{equation}
This is the acoustic waves of cosmology, and in the high frequency limit those obey the standard wave equation in Minkowski space with
$$c_s^2 = \frac{\partial\tilde p}{\partial\tilde \rho} $$
being the sound speed.

If the Reader is not familiar with the cosmological perturbation theory, there is no need of going through all the calculations in detail. The important point is just that the equation (\ref{wave}) for adiabatic waves in the cosmic plasma follows from the gravitational equations alone, and those are simply the GR ones, with the ${\tilde T}_{\mu\nu}$ in the role of energy and momentum quantities. The energy density perturbations $\delta \tilde \rho$, and consequently the Rastall frame ones $\delta \rho$ too, are determined by the Newtonian potentials, therefore it is all about the motion of matter, and its sound speed cannot be equal to $\frac{\partial p}{\partial \rho} $, as opposed to $\frac{\partial\tilde p}{\partial\tilde \rho}$.

In other words, considering both the Rastall gravity equations and the fluid equations of motion with the sound speed of $c_s^2=\frac{\partial p}{\partial\rho}$, whatever they are to be, inevitably leads to contradictions, with practically no perturbations possible at all due to an over-determined system of equations when a field is obliged to obey two wave equations of different velocities simultaneously. In the Ref. \cite{cosmo}, even naively putting the usual energy-momentum tensor of a canonical scalar field in place of the Rastall one did not allow the authors to see non-trivial perturbations.

A couple of important remarks are in order. First note that, when fixing a gauge, I implicitly assumed that the energy-momentum tensor is indeed a tensor under the diffeomorphisms. However, if it was not an invariant object, then the equation $G_{\mu\nu}={\tilde T}_{\mu\nu}$ would be even less consistent, or it would require some restrictions on coordinate choices which then must be explicitly put into the definition of the theory. And anyway, the problem can be understood as the evident fact that the gravitational equations demand that there is a tensor which depends on the matter fields only and has a zero divergence on-shell. And then, if we assume that the matter, following its own equations, behaves according to another energy-momentum tensor, it leads us to contradictions.

Another limitation is that, for sure, we have used above an assumption that there is nothing else in the Universe, except for the ideal fluid which has the acoustic waves in it. One can, of course, try to get non-trivial novel solutions with several fields there \cite{cosmo}. But then it either can also be presented in terms of interacting fields in the standard GR, for the sum of all the energy-momentum tensors ${\tilde T}_{\mu\nu}$ must still be covariantly conserved on the equations of motion of the matter fields, or we face the same problem: adding the gravitational equations would decrease the number of the degrees of freedom in the matter sector due to the contradictions among the equations.

\section{On an action principle for Rastall gravity}

It is often stated in the literature that an action principle for the Rastall gravity is not known. The available attempts \cite{other}, which are closer to the Rastall's initial idea of an effective energy-momentum tensor depending also on curvature, produce some different models, not precisely the Rastall equation. There is no surprise that we were unable to find a natural variational principle for it, since the very thing isn't natural. It is a rewriting of GR which abandons the covariant conservation which directly follows from the fact it is a theory  which is diffeomorphism invariant and has only the metric as its dynamical variable. Therefore, we have to violate one of these conditions, or both as I will do below.

The case of Rastall theory with $l=\frac14$ is that of unimodular gravity. All the other cases can be obtained as a linear combination of the unimodular and the standard GR expressions. The problem is however that the tracelessness of the physical unimodular gravity equation follows from an absolutely arbitrary term proportional to the metric, and this is not what we want in our task. However, it gives us a hint as to what kind of violations we could try.

We define an action via
\begin{equation}
\label{action}
S=- \frac{1}{1+c}\cdot \int d^4 x \left(\sqrt{-g} \left(\vphantom{\sqrt{-g}} R + \zeta (\beta-R)\right) + c \beta  + \lambda \left( \sqrt{-g} -1\right)^2\right)
\end{equation}
with three auxiliary fields $\zeta(x)$, $\beta(x)$, and $\lambda(x)$, and a constant $c$ whose value defines a particluar model. Ignoring the overall prefactor, the variation with respect to the metric gives the equation
\begin{equation}
\label{equation}
(1-\zeta) \cdot G_{\mu\nu} + \left(\bigtriangledown_{\mu}\bigtriangledown_{\nu}- g_{\mu\nu}\square \right)\zeta -\left(\frac12 \zeta\beta + \lambda \left(\sqrt{-g} -1\right)\right)\cdot g_{\mu\nu} =0.
\end{equation}
The $\zeta$ and $\beta$ variations demand that
$$\beta=R \qquad \mathrm{and} \qquad \zeta=-\frac{c}{\sqrt{-g}}$$
respectively. A little less trivial trick is in the $\lambda$ part which yields the unimodular-type condition
$$ \sqrt{-g}=1$$
with an extra term disappearing from the equation of motion (\ref{equation}) due to quadratic vanishing of the corresponding Lagrangian term. These conditions turn the equation (\ref{equation}) into
$$(1+c) \cdot G_{\mu\nu}+ \frac{c}{2} R g_{\mu\nu} =0.$$
Restoring the $\frac{1}{1+c}$ prefactor, we see that variation of the action (\ref{action}) directly produces the equation
$$R_{\mu\nu} + \left(\frac{c}{2(1+c)} - \frac12\right)\cdot R g_{\mu\nu} =0$$
which is the Rastall one (\ref{Rastall}) with $l=\frac{c}{2(1+c)}$. 

I must admit that, at the face value, this approach restricts the freedom of choosing the coordinates, due to the $\sqrt{-g}=1$ condition. However, one can covariantise it by standard methods of adding even more structures to the action \cite{unimod}. One more potential problem is the singular character of the variational principle as I have used quadratic  vanishing of a term in the action at the solution of $\sqrt{-g}=1$. In modern theoretical physics, even crazier structures are sometimes used \cite{Vikman}, though they always look very worrisome \cite{me}. However, an action principle is not my main topic for this paper.

Another question to ask is how to add matter to the model. After all, this theory in vacuum results in the same $R_{\mu\nu}=0$ equation as GR. As was also correctly pointed out in the Ref. \cite{Visser}, one can look at an action principle for Rastall gravity as performing some unnatural separation of the general relativistic action into the gravity and the matter parts. Alternatively, what I have done above is producing the left hand side of the Rastall equation (\ref{Rastall}) from a pure gravity action functional. However then one must also generate the Rastall non-conserved energy-momentum tensor of matter from its own part of the action which, in turn, isn't a trivial task either. 

Taking a standard, diffeo-invariant action of, say, a canonical scalar field generically leads to an over-determined system of equations by producing the energy-momentum tensor which better corresponds to the Einstein equation than to the Rastall one. More precisely, taking the standard conserved tensor in the role of Rastall $T_{\mu\nu}$ requires, by virtue of the formula (\ref{nuconsv}), that the energy-momentum trace be constant. A possible way to go is to perform the same trick with the matter Langrangian density ${\mathcal L}_m$ as has been done with the gravitational $R$ term above (\ref{action}), so that after having integrated the auxiliary fields out, the combination (\ref{back}) of the ${\mathcal L}_m$ variations would appear in the gravitational equations of motion.

\section{Conclusions}

The motivation of Rastall's work \cite{Rastall} was in coupling the matter fields to curvature. No doubt, it is an interesting idea in itself. However, the final result (\ref{Rastall}) has nothing to do with it. There is nothing in it but a simple redefinition (\ref{back}) of what we call the energy-momentum tensor. This is the claim of Visser \cite{Visser} which I fully support. There were some refutations of it though \cite{antiV}. And indeed, one can always say that it is the effective energy-momentum tensor, what the Einstein tensor is equal to, for any modified gravity model. But then the model remains undetermined, unless we give a precise shape of this new entity. It's not the case of Rastall gravity when $T_{\mu\nu}$ is assumed to be a characteristic of matter.

In the Rastall gravity, we need that the energy-momentum tensor satisfies an unusual generalisation of covariant conservation (\ref{nuconsv}) on-shell, for otherwise the gravity and the matter equations would be contradictory to each other, pathologically decreasing the number of dynamical degrees of freedom in the matter fields when coupling them to gravity. It means that the tensor ${\tilde T}_{\mu\nu}$ is covariantly conserved, and can be obtained in the usual way from the usual actions. At the same time, the new tensor $T_{\mu\nu}$ does not correpond to any conservation law, not even in Minkowski. It makes it very unnatural, to expect then that the physical matter properties should correspond to naive usage of the $T_{\mu\nu}$ quantities in some textbook formulae. Moreover, as I've shown above, it can be just inconsistent.

All in all, my conclusion is that Rastall gravity, despite the initial motivations of a clearly gravitational origin, is precisely GR with an algebraic redefinition of the energy-momentum tensor. In case we assume that the Rastall tensor $T_{\mu\nu}$ is somehow more physical than the standard ${\tilde T}_{\mu\nu}$, it is more about unusual properties of matter fields than gravity. Moreover, a naive idea like that would often lead to inconsistencies, too.

\end{document}